\documentclass[a4paper,11pt]{article}
\usepackage{pos}
\title{Search for dark matter around intermediate mass black holes with the H.E.S.S. experiment}

\author[a]{J. Aschersleben}
\author[b,*]{D. Horns }
\author[c,*]{E. Moulin}
\author[a,*]{M. Vecchi, \note{Corresponding author.}}
\author{for the H.E.S.S. Collaboration}

\affiliation[a]{Kapteyn Astronomical Institute, University of Groningen, PO Box 800, NL-9700 AV, Groningen, The Netherlands}
\affiliation[b]{Institut für Experimentalphysik, University of Hamburg, Luruper Chaussee 149, 22761 Hamburg, Germany}
\affiliation[c]{IRFU, CEA, Université Paris-Saclay, F-91191 Gif-sur-Yvette, France}

\emailAdd{m.vecchi@rug.nl}

\abstract{Intermediate mass black holes (IMBHs), with masses ranging from a hundred and a million solar masses, are hypothesised to be surrounded by dense regions of dark matter known as dark matter spikes, where the annihilation of dark matter particles could produce detectable gamma rays. The detection of dark matter annihilation around IMBHs therefore offers a promising approach for probing the nature of dark matter. In this work, we search for dark matter annihilation around IMBHs using data from the Galactic Plane Survey, the Extragalactic Survey and a selection of satellite galaxies observed by the H.E.S.S. gamma-ray experiment in Namibia. Since no evidence for a gamma-ray signal from dark matter annihilation around IMBHs has been found, we set upper limits on the velocity-weighted annihilation cross section for dark matter masses between 800 GeV and 100 TeV. Our analysis obtains limits on the velocity-weighted annihilation cross section below the thermal relic cross section for dark matter masses between 10 and 100 TeV.}

\ConferenceLogo{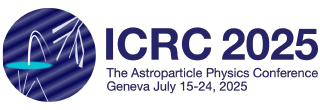}

\FullConference{39th International Cosmic Ray Conference (ICRC2025)\\
 15–24 July 2025\\
Geneva, Switzerland\\}

\begin{document}
\maketitle

\section{Introduction}
The characteristics and composition of dark matter (DM) remain among the most significant unresolved issues in fundamental physics~\cite{Bertone:2004pz}. Evidence for its existence comes from various astrophysical observations~\cite{DMev}, including the rotational curve of spiral galaxies, the behaviour of galaxy clusters, the cosmic microwave background, and the large-scale structure of the universe. Leading DM candidates include Weakly Interacting Massive Particles (WIMPs)~\cite{WIMP}. These particles are predicted by theories beyond the Standard Model of particle physics and are thought to have masses ranging from GeV to 100 TeV. WIMPs are expected to produce gamma rays through self-annihilation processes~\cite{IDM}, offering a potential method for indirectly detecting DM via high-energy neutrinos, gravitational waves, cosmic rays and gamma rays. Ground-based gamma-ray observatories are used to detect TeV gamma rays and thereby search for WIMP DM targeting a variety of astrophysical objects, which are known to be DM rich.
Intermediate-mass black holes (IMBHs)~\cite{IMBHrev} emerge as a new class of targets for indirect DM searches with gamma rays. They are generally defined as having masses between approximately $10^2$ and $10^6$~$M_\odot\ $, filling the gap between stellar mass black holes and supermassive black holes (SMBHs).
The first unambiguous evidence for the existence of IMBHs comes from the gravitational wave event GW190521~\cite{LIGOScientific:2020}, which was the result of a binary black hole merger event which generated a black hole with a mass of 
about 142$M_\odot\ $. The DM spike model initially developed for the supermassive black hole (SMBH) located at the centre of our galaxy, Sagittarius A*~\cite{Gondolo:1999ef}, was 
 extended to IMBHs in~\cite{Bertone:2005}.


\section{Expected signal}
\noindent The DM density profile around IMBHs and the corresponding gamma-ray flux from DM self annihilation can be computed using the approach initially proposed in~\cite{Gondolo:1999ef} 
and later refined in \cite{Bertone:2005,Aschersleben:2024xsb}. 

\begin{figure}
    \centering    \includegraphics{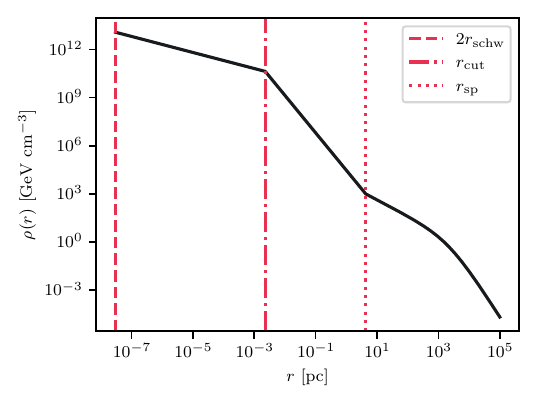}
    \caption{Dark matter density profile around an IMBH assuming a black hole mass of $1.6~10^5  M_{\odot}$. Figure taken from~\cite{Aschersleben:2024xsb}.}
    \label{fig:spike_profile}
\end{figure}
\noindent A schematic description of the DM density profile around an IMBH is given in figure~\ref{fig:spike_profile}. Four regions are clearly visible. At large radii the density profile follows the one of the host halo where the black hole has formed. The DM density distribution is assumed to follow a Navarro, Frenk and White (NFW) profile~\cite{Navarro:1996}. The DM density $\rho_\mathrm{sp}(r)$ within the spike region $r \leq r_\mathrm{sp}$, where $r$ indicates the distance to the centre of the IMBH, is described by a power-law given by
\begin{equation}\label{eq:spike}
    \rho_{\mathrm{sp}}(r) = \rho_\mathrm{NFW}(r_{\mathrm{sp}}) \left(\frac{r}{r_{\mathrm{sp}}}\right)^{-\gamma_{\mathrm{sp}}} \, ,
\end{equation}
where $\rho_\mathrm{NFW}$~~\cite{Navarro:1996} is the NFW profile, $r_\mathrm{sp}$ the spike radius and $\gamma_\mathrm{sp}$ the spike index. The latter depends on the initial power-law index $\gamma$ of the DM density profile of the host halo by $\gamma_{\mathrm{sp}} = (9-2 \gamma)/(4-\gamma)$. Assuming a NFW profile ($\gamma$ = 1),
$\gamma_\mathrm{sp}=7/3$. Towards smaller radii, the so called ``weak cusp" regime is achieved. The spike would bring the density to diverge, but due to the self-annihilation of the DM particles, a saturation effect occurs, leading to the formation of a weak cusp 
 $\rho_\mathrm{wcusp}(r)$, which is given by (assuming \textit{s}-wave annihilation):
\begin{equation}\label{eq:cusp}
    \rho_\mathrm{wcusp}(r) = \rho_\textrm{sp}(r_\mathrm{sat}) \cdot \left( \dfrac{r}{r_\mathrm{sat}} \right)^{-0.5}
\end{equation}
 The density is assumed to be zero 
 as soon as the distance $r$ approaches 2 Schwarzchild radii.

\noindent The gamma-ray flux from DM annihilation
considering the previously described density profile for $2r_\mathrm{schw} \leq r \leq r_\mathrm{sp}$, can be written as~\cite{Aschersleben:2024xsb}:
\begin{align}\label{eq:flux_general}
    \Phi(E,D) & = \left. \frac{1}{2} \frac{\langle \sigma v \rangle}{m_\chi^2} \frac{1}{D^2} \frac{{\rm d}N}{{\rm d}E} \int_{2r_\mathrm{schw}}^{r_{\rm sp}} \rho^2(r) r^2 dr \right. \\
    & \approx \frac{{\rm d}N}{{\rm d}E} \frac{\langle \sigma v \rangle}{m_\chi^2 D^2} \rho (r_\mathrm{sp})^2 r_\mathrm{sp}^3 \frac{2\gamma_\mathrm{sp} - 1}{8\gamma_\mathrm{sp} - 12} \left(  \frac{r_\mathrm{cut}}{r_\mathrm{sp}} \right)^{3 -2 \gamma_\mathrm{sp}} 
    \label{equation:1}
\end{align}
with the DM annihilation spectrum $\mathrm{d}N/\mathrm{d}E$ and the distance $D$ from the observer to the IMBH~\cite{Aschersleben:2024xsb}. 

\noindent The number and spatial distribution of IMBHs in the Milky Way is unknown. We therefore rely on a mock catalogue derived using cosmological simulations~\cite{Aschersleben:2024xsb, zenodo} to compute the DM density profile around IMBHs and the corresponding gamma-ray flux from DM self annihilation.


\noindent The catalogue is based on the EAGLE cosmological simulations~\cite{EAGLE}. In particular, 
about 150 Milky Way-like galaxies are selected at redshift z = 0 based on specific selection criteria~\cite{Aschersleben:2024xsb}. On average, about
15 IMBHs are present within their selection of Milky Way-like galaxies and their associated
satellite galaxies, being mainly distributed towards the Galactic Centre and along
the Galactic Plane. It was also demonstrated that H.E.S.S. should be sensitive enough to
detect VHE gamma-ray signal from DM annihilation around IMBHs for DM masses in the
GeV to TeV range, with a thermal velocity-weighted annihilation cross section.


\section{Analysis of H.E.S.S. data}

\noindent We analyse the observations performed by the  High Energy Stereoscopic System (H.E.S.S.)~\cite{HESS} in the Khomas Highland of Namibia. Since the beginning of operations in 2003, H.E.S.S. has been instrumental in exploring the TeV regime. It is composed of five telescopes: four (CT1-4) have mirror areas of 107~m$^2$, providing a gamma-ray energy threshold of $\sim 100$~GeV, while the fifth (CT5) has a mirror area of 612~m$^2$
and correspondingly lower energy threshold  ($\sim 30$ GeV). H.E.S.S. is the only TeV gamma-ray facility currently operating in the southern hemisphere. 
Thus, it is the best-positioned detector for observations of the Galactic Centre and the surrounding region, allowing for unprecedented sensitivity of the Galactic Plane Survey~\cite{HGPS}. 

\noindent  We analyse H.E.S.S. observations (Total:~$\simeq$ 6200 hrs) from the 
H.E.S.S. Galactic Plane Survey (HGPS)~\cite{HGPS}, the H.E.S.S. Extra Galactic Survey (HEGS)~\cite{HEGS} and the observations of five selected satellite galaxies, namely LMC, SMC, Canis Major, Sagittarius and Fornax~\cite{THISpaper}.


\noindent  Based on EAGLE simulations~\cite{THISpaper}, up to half of the satellite galaxies with stellar mass larger than $2~10^7~M_\odot\ $ contain at least one IMBH.
Only the satellite galaxies visible by H.E.S.S.,  included in~\cite{2012AJ....144....4M} and with  stellar mass larger than $2~10^7~M_\odot\ $ are included in this study. 
The analysis of observations of satellite galaxies is done assuming a (conservative) probability $p_{BH}~=~20\%$ that a satellite galaxy hosts an IMBH.

\subsection{Statistical analysis}
\noindent The process of DM self-annihilation around IMBHs is expected to produce point-like sources of gamma rays. Each of these sources should exhibit identical energy spectra across various IMBHs, with an energy cut-off corresponding to the mass of the DM particle. These sources are likely to remain unidentified, and consequently, our focus is on detecting such signals within observations from the HGPS\cite{HGPS}, HEGS~\cite{HEGS}, and the selected satellite galaxies.

\noindent The HGPS\cite{HGPS} and HEGS~\cite{HEGS} studies provide catalogues of sources, detailing their coordinates, spatial extensions, and energy spectra. Previous analyses using H.E.S.S. have thoroughly examined the selected satellite galaxies, including the LMC, SMC, Canis Major, Sagittarius, and Fornax, finding no evidence for gamma-ray sources. For our current work, the analysis comprises two primary steps: first, identifying potential candidates for DM annihilation around IMBHs by searching for unidentified, point-like sources within the HGPS and HEGS catalogues~\cite{HEGS} and the satellite galaxy analyses; and second, if no IMBH candidates are identified, establishing upper limits on the velocity-averaged DM cross-section.

\begin{figure}
    \centering
\includegraphics{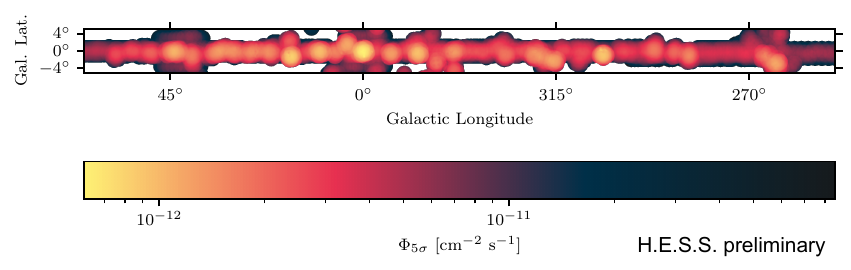}
    \caption{Flux sensitivity map for the HGPS in Galactic Coordinates for a DM mass of $m_\chi = 30~TeV$, the $b \overline{b}$-channel.}
    \label{fig:sensitivity}
\end{figure}

\noindent We aim at determining if H.E.S.S. is sensitive enough to detect gamma-ray fluxes from WIMP annihilation around IMBHs. 
To calculate the flux sensitivity maps, we use version 1.2 of the Python package Gammapy~\cite{gammapy} and adhere to the methodology outlined in~\cite{HGPS}. 
With respect to the published analyses, we recompute the flux sensitivity for all DM masses, and all FoVs, considering the 
$b \overline{b}$ annihilation channel as our benchmark. 
In addition, we use the DM annihilation spectra from~\cite{Cirelli} instead of the standard power law spectrum. An example is given in Fig.~\ref{fig:sensitivity}, showing a flux sensitivity map for DM mass of 30 TeV and the $b \overline{b}$ annihilation channel.


\noindent The HEGS~\cite{HEGS} source catalogue and observations of satellite galaxies did not reveal any unidentified point-like sources. However, the HGPS source catalogue~\cite{HGPS} includes three such sources: HESS J1741-302, HESS J1745-290, and HESS J1746-285. While HESS J1741-302, and HESS J1745-290 do not show any prominent cut-off feature in their spectra,
the spectrum of HESS J1746-285 displays a significant energy cut-off, making it an interesting candidate for our study. However, plausible astrophysical (non-DM) explanations for its TeV emission are available. 
Therefore, we conclude that our search did not yield any unambiguous IMBH candidate and we therefore set upper limits on the velocity-averaged DM cross-section as a function of the DM mass. 
For each IMBH in the catalogue that is not linked to a satellite galaxy, we initially verified if it lies within the observational field of view (FoV) of the HGPS and HEGS surveys. We then computed the anticipated gamma-ray flux from DM annihilation, denoted as $\Phi_\chi$, for specific DM masses ranging from 800~GeV to 100~TeV and corresponding cross-sections. This calculated flux was compared against the flux sensitivity, $\Phi_{5 \sigma}$, derived from sensitivity maps at the IMBH's location to assess detectability. IMBHs situated in exclusion zones were deemed undetectable.


\noindent We compute the upper limits using the methodology described in~\cite{HESS:2008}, which describes the results of a similar analysis using a much smaller dataset, focused on the Galactic Plane, and a catalogue based on~\cite{Bertone:2005}. The IMBHs identified as detectable from the analysis of satellite galaxies are merged with those detected by the HGPS and HEGS. The distribution of detectable IMBHs within Milky Way-like galaxies, is then computed for each DM mass, for annihilation into $b \overline{b}$. Additional channels are under study~\cite{THISpaper}.

\begin{figure}[htb]
    \centering    \includegraphics{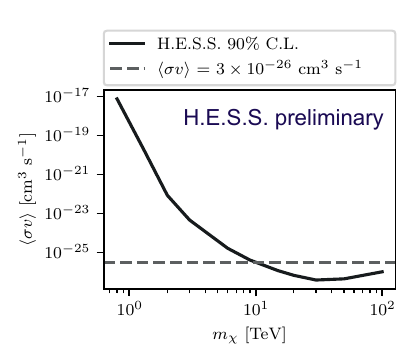}
    \caption{Upper limits on velocity averaged DM annihilation cross section $\langle \sigma v \rangle$ as a function of DM mass $m_\chi$ on a $90 \%$ C.L. The grey dashed line corresponds to the thermal relic cross section of $\langle \sigma v \rangle = 3~10^{-26}~cm^{-3}s^{-1}$, for the $b \overline{b}$ annihilation channel.}
    \label{fig:dm_limits}
\end{figure}

\noindent The distribution of Milky Way-like galaxies containing intermediate-mass black holes (IMBHs) can be effectively modelled using a Poisson distribution \( P(k, \lambda) \). We fit a Poisson distribution to the histogram of detectable IMBH counts to determine the best-fit parameter \(\tilde{\lambda}\). Using this parameter, we calculated the probability of not detecting any IMBHs and subsequently set upper limits on \(\langle \sigma v \rangle\) at a 90\% confidence level. Specifically, we determined the value of \(\langle \sigma v \rangle\) that ensures a 90\% probability of detecting at least one IMBH, which translates to \( P(k \geq 1, \tilde{\lambda}) = 0.9 \). Consequently, this results in a 10\% probability of detecting no IMBHs, represented as \( P(k = 0, \tilde{\lambda}) = 0.1 \).


\section{Results}
\noindent Our results are shown in Fig.\ref{fig:dm_limits} for the $b \overline{b}$ annihilation channel. The curve exhibits a downward trend as the DM mass increases. Beyond 10 TeV our exclusion curve falls below the thermal relic cross-section, with the most stringent constraint reaching a value of $4~10^{-27} cm^{-3} s^{-1}$ at a DM mass of 30 TeV. These results do not include systematic uncertainties and their impact is under investigation~\cite{THISpaper}. 

\section{Conclusions}
\noindent We explored the possibility of detecting DM annihilation around IMBHs based on the expected source distribution from a mock catalogue
and analysing existing observations of the H.E.S.S. Galactic Plane Survey (HGPS)~\cite{HEGS}, of the H.E.S.S. Extra Galactic Sky survey (HEGS)~\cite{HEGS} and the observations of selected satellite galaxies, corresponding to a total of 6200 hours.

\noindent We found no unambiguous evidence of an  point-like source signal with spectral features indicative of DM annihilation in IMBHs, and we therefore established upper limits on the velocity-weighted DM annihilation cross-section,  for DM masses between 800 GeV and 100 TeV. Our findings constrain the thermal relic cross-section hypothesis for DM masses higher than 10 TeV. These limits are subject to  systematic uncertainties related to the formation and evolution of IMBHs and their associated DM spikes.

\end{document}